%% file: main.tex
\documentclass[sigplan,screen]{acmart}
\input{preamble}

\AtBeginDocument{%
  }

\setcopyright{cc}
\setcctype{by}
\acmDOI{10.1145/3837729.3840480}
\acmYear{2026}
\copyrightyear{2026}
\acmISBN{979-8-4007-2910-2/2026/10}
\acmConference[SPLASH Companion '26]{Companion Proceedings of the 2026 ACM SIGPLAN International Conference on Systems, Programming, Languages, and Applications: Software for Humanity}{October 3--9, 2026}{Oakland, CA, USA}
\acmBooktitle{Companion Proceedings of the 2026 ACM SIGPLAN International Conference on Systems, Programming, Languages, and Applications: Software for Humanity (SPLASH Companion '26), October 3--9, 2026, Oakland, CA, USA}
\acmSubmissionID{splashcomp26demo-p3-p}
\received{2026-06-27}
\received[accepted]{2026-07-25}

\begin{document}

\title{FlowLog: Re-thinking Datalog for Fast and Extensible Static Analysis}

\author{Zhenghong Yu}
\correspondingauthor
\orcid{0009-0000-0841-0694}
\affiliation{%
  \institution{University of Wisconsin-Madison}
  \city{Madison}
  \country{USA}
}
\email{m0riarty@cs.wisc.edu}

\author{Hangdong Zhao}
\orcid{0009-0009-7636-0831}
\affiliation{%
  \institution{Microsoft Gray Systems Lab}
  \city{Redmond}
  \country{USA}
}
\email{hangdongzhao@microsoft.com}

\author{Wanzhu Hou}
\orcid{0009-0001-5146-3289}
\affiliation{%
  \institution{University of Wisconsin-Madison}
  \city{Madison}
  \country{USA}
}
\email{whou25@wisc.edu}

\author{Paraschos Koutris}
\orcid{0000-0001-6309-1702}
\affiliation{%
  \institution{University of Wisconsin-Madison}
  \city{Madison}
  \country{USA}
}
\email{paris@cs.wisc.edu}

\renewcommand{\shortauthors}{Zhenghong Yu, Hangdong Zhao, Wanzhu Hou, and Paraschos Koutris}

\input{sections/frontmatter}

\maketitle

\input{sections/introduction}
\input{sections/tool-overview-and-usage}
\input{sections/evaluation}
\input{sections/extensibility}
\input{sections/tool-availability}


\bibliographystyle{ACM-Reference-Format}
\bibliography{references}

\end{document}

%% file: preamble.tex
\usepackage[utf8]{inputenc}

\usepackage{microtype}
\usepackage{siunitx}

\definecolor{cmdgray}{gray}{0.20}
\definecolor{outgray}{gray}{0.45}
\definecolor{argrelc}{RGB}{30,110,110}
\definecolor{argtuplec}{RGB}{90,110,30}
\definecolor{argdiffc}{RGB}{120,60,120}

\usepackage{subcaption}

\usepackage{listings}

\lstdefinelanguage{Rust}{
  morekeywords={
    abstract,alignof,as,become,box,break,const,continue,crate,do,else,enum,
    extern,false,final,fn,for,if,impl,in,let,loop,macro,match,mod,move,mut,
    offsetof,override,priv,proc,pub,pure,ref,return,self,Self,sizeof,static,
    struct,super,trait,true,type,typeof,unsafe,unsized,use,virtual,where,while,yield,
  },
  morekeywords=[2]{
    arrange, map, filter, flat_map, inspect, distinct, concat, reduce,
    join, join_map, join_core, semijoin, iterate
  },
  sensitive=true,
  morecomment=[l]{//},
  morecomment=[s]{/*}{*/},
  morestring=[b]{"}
}

\lstdefinelanguage{Datalog}{
  sensitive=true,
  morecomment=[l]{//},
  morestring=[b]{"}
}

\lstdefinelanguage{FlowLogShell}{
  morekeywords={begin,txn,put,file,commit,done,abort,rollback,help,h,quit,exit,q},
  sensitive=true
}

\lstdefinestyle{DatalogStyle}{
  language=Datalog,
  basicstyle=\scriptsize\ttfamily,
  columns=fullflexible,
  keepspaces=true,
  showstringspaces=false,
  commentstyle=\color[rgb]{0,0.5,0},
  frame=none,
  aboveskip=4pt,
  belowskip=4pt,
}

\lstdefinestyle{FlowLogShellAuto}{
  language=FlowLogShell,
  basicstyle=\scriptsize\ttfamily,
  columns=fullflexible,
  keepspaces=true,
  showstringspaces=false,
  breaklines=true,
  breakatwhitespace=true,
  frame=none,
  numbers=none,
  xleftmargin=0pt,
  xrightmargin=0pt,
  keywordstyle=\color{cmdgray}\bfseries,
  escapeinside={(@}{@)},
  literate=
    {>>}{{{\color{cmdgray}\bfseries\ttfamily >>}}}2
}

\usepackage{xspace}

\newcommand*\circledwhite[1]{%
  \raisebox{-0.08ex}{\scalebox{1.15}{%
    \textcolor{gray}{\ifcase#1\or\ding{182}\or\ding{183}\or\ding{184}\fi}}}}

\newcommand{\Datalog}{\text{Datalog}\xspace}

\newcommand{\IDB}{\textsc{idb}\xspace}

\newcommand{\EDB}{\textsc{edb}\xspace}

\newcommand{\ir}{IR\xspace}

\newcommand{\flowlog}{\textsf{FlowLog}\xspace}
\newcommand{\doop}{\textsc{DOOP}\xspace}
\newcommand{\polonius}{\textsc{Polonius}\xspace}
\newcommand{\souffle}{\text{Soufflé}\xspace}
\newcommand{\ddlog}{\text{DDlog}\xspace}

\newcommand{\shout}[1]{\textcolor{outgray}{\ttfamily #1}}
\newcommand{\argrel}[1]{\textcolor{argrelc}{\ttfamily #1}}
\newcommand{\argtuple}[1]{\textcolor{argtuplec}{\ttfamily #1}}
\newcommand{\argdiff}[1]{\textcolor{argdiffc}{\ttfamily #1}}

%% file: sections/frontmatter.tex

\begin{abstract}
Datalog is widely used to build static analyzers, yet existing engines often force a tradeoff between efficiency and extensibility.  In practice, static analyses are not run once and forgotten: users edit facts, tune rules, diagnose bottlenecks, and often need semantics beyond standard Datalog, leaving these tasks to ad hoc tooling or invasive engine rewrites.

We demonstrate \textsf{FlowLog}, a Datalog compiler that turns Soufflé-style programs into Differential Dataflow executables for efficient and extensible static analysis. Across 24 benchmarks derived from real-world workloads, \textsf{FlowLog} consistently outperforms state-of-the-art engines in runtime while remaining memory-efficient and scaling better.

The demonstration uses a \textsc{DOOP} points-to analysis.  Attendees \emph{run} it, switching the same program from one-shot to incremental evaluation that retracts a fact and updates results in milliseconds; \emph{tune} it, inspecting per-operator costs in a browser-based profiler and repairing a bad join order; and \emph{extend} it with a k-core example beyond standard Datalog.
\end{abstract}

\ccsdesc[500]{Software and its engineering~Automated static analysis}
\ccsdesc[500]{Software and its engineering~Compilers}

\keywords{Datalog, static analysis tools, efficient execution, extensible
  systems, incremental computation, profiling}

%% file: sections/introduction.tex

\section{Introduction}
\label{sec:intro}

\Datalog is a powerful declarative language for static analysis.  Users describe relations and rules; the engine handles joins, propagation, and fixpoint iteration.  This abstraction has been widely applied to many workloads such as points-to analysis, call-graph construction, borrow checking, and binary disassembly~\cite{doop,polonius,ddisasm}.

This success has inspired the development of \Datalog engines. Systems such as \souffle~\cite{scholz2016fast}, Flix~\cite{flix}, Flan~\cite{flan}, and Ascent~\cite{Ascent} achieve high performance through domain-specific optimizations. While effective in their target settings, these systems tend to sacrifice extensibility: adding incremental maintenance or new semantics typically requires substantial system changes, and scaling to larger workloads frequently demands extensive engineering.  Other systems avoid building from scratch by layering \Datalog on top of existing databases or dataflow frameworks. RecStep~\cite{recstep} compiles \Datalog into SQL and evaluates it one iteration at a time, but switching to the database on each iteration adds high synchronization overhead.  DDlog~\cite{ddlog} compiles \Datalog directly into Differential Dataflow~\cite{DD} (\textsc{DD}), yet empirical studies report substantial memory overhead, sometimes orders of magnitude higher than alternatives~\cite{DBLP:conf/pldi/HuZJS21, DBLP:journals/pacmpl/LiSZ22}.

\flowlog~\cite{zhao2025flowlog} addresses this tradeoff with a new design that pursues efficiency and extensibility together.  For efficiency, it builds on \textsf{DD}'s off-the-shelf streaming operators as execution primitives.  For flexibility, it introduces a relational intermediate representation (\ir) that separates recursive control from each rule's logical plan, opening a dedicated layer for Datalog-aware rewrites that improve one-shot performance while enabling extensions such as incremental maintenance, profiling, and extended semantics.

This paper presents a tool demonstration of \flowlog as a user-facing artifact. Specifically, it makes the following contributions:
\begin{enumerate}
  \item \textbf{Tool overview and end-to-end usage.} Section~\ref{sec:overview} presents \flowlog's system overview and shows how \souffle-style \Datalog programs compile to standalone parallel executables.
  \item \textbf{Efficiency validation.} Section~\ref{sec:efficiency-eval} evaluates \flowlog on 24 real static-analysis workloads, including \polonius-style borrow checking and \doop-based points-to analysis, against modern \Datalog engines.
  \item \textbf{Extensible workflow.} Section~\ref{sec:ext} demonstrates incremental maintenance, profiling, and extended semantics through runnable examples.
\end{enumerate}

The intended users are \emph{software analysis researchers} evaluating new declarative analyses, \emph{compiler and verification tool builders} who embed recursive analyses into larger tools, and \emph{engineers} who operate deployed \Datalog analyses and tune their performance.

A technical paper describes \flowlog's compiler design, optimization techniques, and comprehensive evaluation in detail~\cite{zhao2025flowlog}.

%% file: sections/tool-overview-and-usage.tex

\section{Tool Overview and Usage}
\label{sec:overview}

\begin{figure}[t]
  \centering
  \includegraphics[width=0.8\columnwidth]{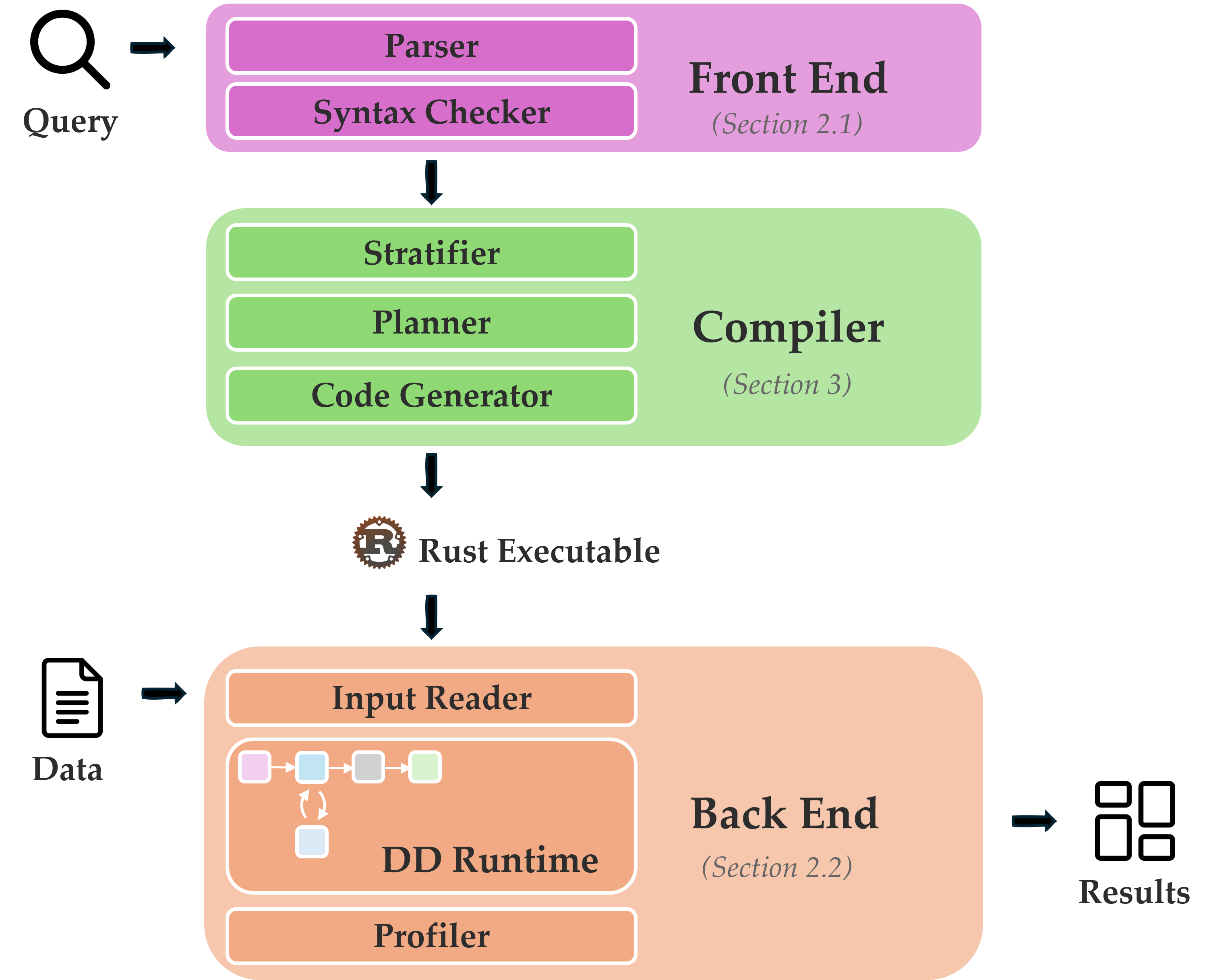}
  \caption{\flowlog architecture.}
  \Description{FlowLog architecture with a front end, compiler, and Differential Dataflow back end.}
  \label{fig:arch}
\end{figure}

As in Fig.~\ref{fig:arch}, \flowlog proceeds in three stages: \circledwhite{1}~\textbf{front end} accepts a \Datalog program in \souffle-style~\cite{scholz2016fast} syntax; \circledwhite{2}~\textbf{compiler} stratifies the program, plans each rule, and generates an executable; and \circledwhite{3}~\textbf{back end} reads the input data, runs the generated program on top of \textsf{DD} to produce the query output, and lets the profiler expose runtime metrics collected during execution.

\subsection{Front End}

\flowlog's front-end accepts \Datalog programs in a \souffle-style syntax, with extensions such as richer types and recursive aggregation. After syntax and type checking, the front-end passes the validated program to the compiler.

We use the \doop points-to analysis as the running example for compilation, execution, and profiling.

\begin{example}[\doop Points-to]\label{ex:doop}
The program below shows part of the core logic of the \doop points-to analysis for Java programs.  For brevity, we elide many declarations and rules; the full program is included in the artifact.
\begin{lstlisting}[style=DatalogStyle, numbers=none, basicstyle=\scriptsize\ttfamily]
// Input (EDB) relations extracted from Java bytecode
// Methods reachable from the program entry points
.decl Reachable(method: str)
// A field load in method inmethod: to = base.sig
.decl LoadInstanceField(base: str, sig: str, to: str, inmethod: str)
// ... other input relations omitted

// Derived (IDB) relations
// Variable var may point to object heap
.decl VarPointsTo(heap: str, var: str)
// Field sig of object baseHeap may point to object heap
.decl InstanceFieldPointsTo(heap: str, sig: str, baseHeap: str)
// ... other derived relations omitted

// Propagate points-to through a field load
VarPointsTo(heap, to) :-
  Reachable(inmethod),
  LoadInstanceField(base, sig, to, inmethod),
  VarPointsTo(baseheap, base),
  InstanceFieldPointsTo(heap, sig, baseheap).
// bad join order for the profiler demo: .plan (1, 3, 4, 2)
// ... other rules omitted

.output VarPointsTo
\end{lstlisting}
\end{example}

\subsection{Compiler}

\flowlog compiles a validated \Datalog program into a standalone executable.
The same source program can be compiled for one-shot execution, incremental
maintenance, or profiling. Table~\ref{tab:flags} lists
the main options; the command below compiles Example~\ref{ex:doop} in one-shot mode:
\begin{lstlisting}[numbers=none, basicstyle=\scriptsize\ttfamily, columns=fullflexible]
$ flowlog doop.dl -o doop-one-shot -F ./facts -D ./out
\end{lstlisting}
\begin{table}[htbp]
  \centering
  \caption{\flowlog compile options.}
  \label{tab:flags}
  \footnotesize
  \begin{tabular}{@{}l@{\hspace{1.5em}}l@{}}
    \toprule
    Option & Description \\
    \midrule
    \texttt{-D <dir>}     & output result directory \\
    \texttt{-F <dir>}     & input fact directory \\
    \texttt{--mode <m>}   & execution mode; one-shot (default) or incremental \\
    \texttt{-o <path>}    & output binary \\
    \texttt{-P}           & enable profiling \\
    \bottomrule
  \end{tabular}
\end{table}

\begin{figure}[!htbp]
  \centering
  \includegraphics[width=\columnwidth]{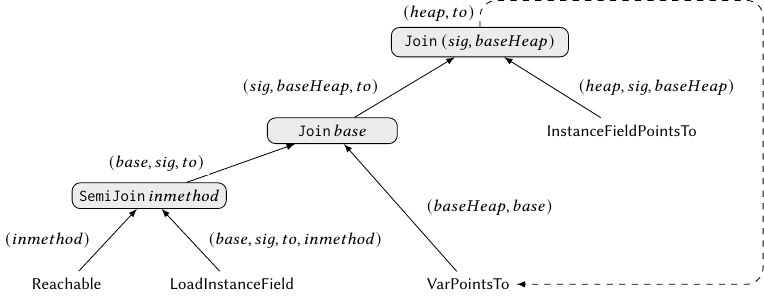}
  \caption{Optimized \ir for the rule in Example~\ref{ex:doop}.}
  \Description{IR tree for the points-to field-load rule: a Reachable semijoin filters the loads, two fused Join operators perform the joins, and a dashed edge feeds the results back into the recursion.}
  \label{fig:ir}
\end{figure}

\begin{sloppypar}
Internally, the compiler stratifies the program and lowers each rule into a relational intermediate representation (\ir).  The \ir is a tree of operators: leaves are input relations, and edges carry tuple schemas.  Figure~\ref{fig:ir} shows the \ir for the rule in Example~\ref{ex:doop}.  A \texttt{SemiJoin} first restricts \texttt{LoadInstanceField} to methods in \texttt{Reachable}, and two \texttt{Join} operators then combine the result with \texttt{VarPointsTo} and \texttt{InstanceFieldPointsTo} into new \texttt{VarPointsTo} tuples.  The dashed edge feeds these tuples back into the recursion.
\end{sloppypar}

Join order largely determines intermediate result sizes, so the planner applies several \Datalog-aware optimizations to shrink them and make execution less sensitive to join order: \circledwhite{1} \emph{pushdown}, which moves projections, filters, and semi/anti-joins as early as possible; \circledwhite{2} \emph{sideways information passing}, a Yannakakis-style semijoin reduction~\cite{Yannakakis81} that pre-filters atoms before the actual joins; and \circledwhite{3} \emph{subplan sharing}, which reuses repeated indexes across the program.

The code generator then emits a \textsf{DD} dataflow in Rust from the optimized \ir and builds the standalone executable.  Each \ir operator maps to one or more \textsf{DD} operators, and \flowlog relies on \textsf{DD} to run the program incrementally and in parallel to a fixpoint.  The generated code also includes profiling instrumentation, discussed in Sec.~\ref{sec:ext}.

\subsection{Back End}

The back end loads the \EDB facts, executes the binary on the \textsf{DD} runtime to a fixpoint, and derives the \IDB results. The resulting binary can then be run with:

\begin{lstlisting}[style=DatalogStyle, numbers=none, basicstyle=\scriptsize\ttfamily, columns=fullflexible]
$ ./doop-one-shot -w 32    // run in parallel with 32 workers
\end{lstlisting}

\textsf{DD} is a streaming dataflow library.  Its relations are collections of tuple updates tagged with logical time and an integer multiplicity, so operators process changes rather than whole relations.  This yields efficient semi-naive evaluation: recursive iterations process only newly derived tuples, run in parallel across workers, and keep indexed state in arrangements that the profiler can inspect.

%% file: sections/evaluation.tex

\section{Efficiency Evaluation}
\label{sec:efficiency-eval}

\begin{figure*}[t]
  \centering
  \includegraphics[width=\textwidth]{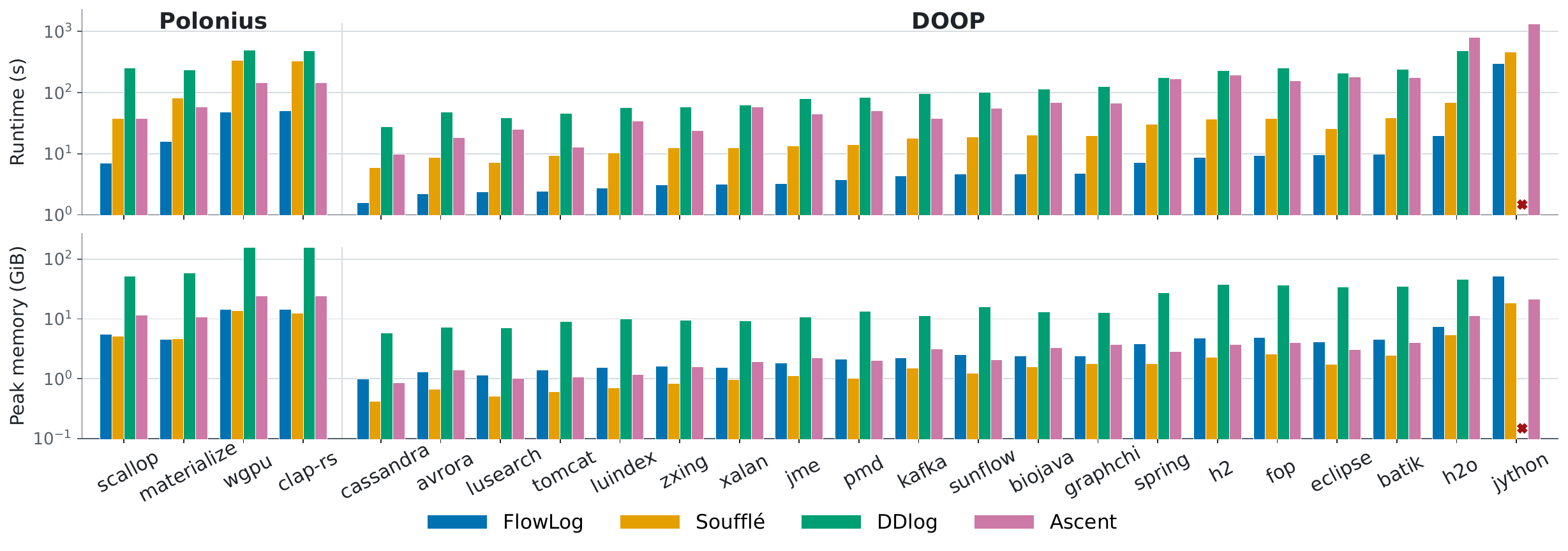}
  \caption{Runtimes (top) and peak memory (bottom) of \flowlog, \souffle, \ddlog, and Ascent on \polonius borrow-checking (left) and \doop points-to (right) workloads for 32 threads (log scale; lower is better).  \textcolor{red!60!black}{\Large\textbf{×}} indicates a timeout.  \flowlog is the fastest on every benchmark in our study and uses far less memory than \ddlog.}
  \Description{Runtime and peak-memory comparison of FlowLog, Souffle, DDlog, and Ascent across Polonius and Doop workloads.}
  \label{fig:comparison}
\end{figure*}

\subsection{Evaluation Setup}

\subsubsection{Workloads} We use two real static-analysis workloads.  \emph{\doop}~\cite{doop} is a Java points-to analysis; we run a core, string-keyed port of it on real-world Java programs from the DaCapo suite (e.g., \texttt{tomcat}, \texttt{eclipse}).  \emph{\polonius}~\cite{polonius} is the Rust borrow checker; we run a simplified, integer-keyed version on Rust crates (e.g., \texttt{wgpu}, \texttt{clap-rs}).

\subsubsection{Competing Engines} We compare against several state-of-the-art Datalog engines: (1) \textbf{\souffle} (compiled) with its mature optimizer~\cite{DBLP:conf/pldi/HuZJS21, IndexSouffle, TrieSouffle, DBLP:conf/lopstr/ArchHZSS22}; (2) \textbf{\ddlog}~\cite{ddlog}, which shares the same \textsf{DD} backend as \flowlog but lacks its relational \ir and optimizations, making it a fair baseline. We also report results from (3) \textbf{Ascent}~\cite{Ascent}, a \Datalog embedded in Rust through compile-time macros.  We omit other \Datalog systems because they are either not open source or lack support for features our workloads require. Reported runtimes exclude compilation time. During timing runs, programs emit only output cardinalities to avoid output-I/O overhead. We separately validated that all engines produce the same results on each benchmark.

\subsubsection{Environment} The benchmark artifact records the engine versions and scripts used for runtime, memory, and scalability measurements.\footnote{\url{https://github.com/flowlog-rs/flowlog-bench}} Experiments run on a CloudLab virtual machine~\cite{cloudlab} with two AMD EPYC 7543 32-core processors, hyper-threading enabled, and \SI{256}{\giga\byte} RAM.

\subsection{Results Summary}

\subsubsection{Runtime} Figure~\ref{fig:comparison} (top) reports execution time on a log scale.  \flowlog is the fastest on all 24 benchmarks.  On \doop, its median speedup is $3.9\times$ over \souffle ($1.5$--$4.3\times$), $22\times$ over \ddlog ($16$--$27\times$), and $14\times$ over Ascent ($4.4$--$41\times$); on \polonius, it is $5.9\times$ over \souffle ($5.1$--$7.0\times$), $13\times$ over \ddlog ($9.5$--$35\times$), and $3.3\times$ over Ascent ($2.9$--$5.3\times$).  \flowlog's \ir-level optimizations (Section~\ref{sec:overview}) shrink and reuse intermediate state, and \textsf{DD}'s asynchronous runtime keeps every operator active across the iterations.

\subsubsection{Memory} Figure~\ref{fig:comparison} (bottom) reports peak memory.  \flowlog is far more memory-efficient than \ddlog, which uses a median of $6.3\times$ more ($5$--$13\times$), because \flowlog's subplan sharing avoids the redundant intermediate state that \ddlog materializes.  Against \souffle, \flowlog uses about $2\times$ more memory on \doop ($1.3$--$2.8\times$) and stays on par on \polonius ($\sim1.1\times$).  This is a deliberate trade-off: \flowlog materializes all of its indexes as \textsf{DD} arrangements, which speeds execution and, unlike \souffle's one-shot-only design, leaves it ready to maintain results incrementally (Section~\ref{sec:ext}).

\begin{figure}[htbp!]
  \centering
  \includegraphics[width=\columnwidth]{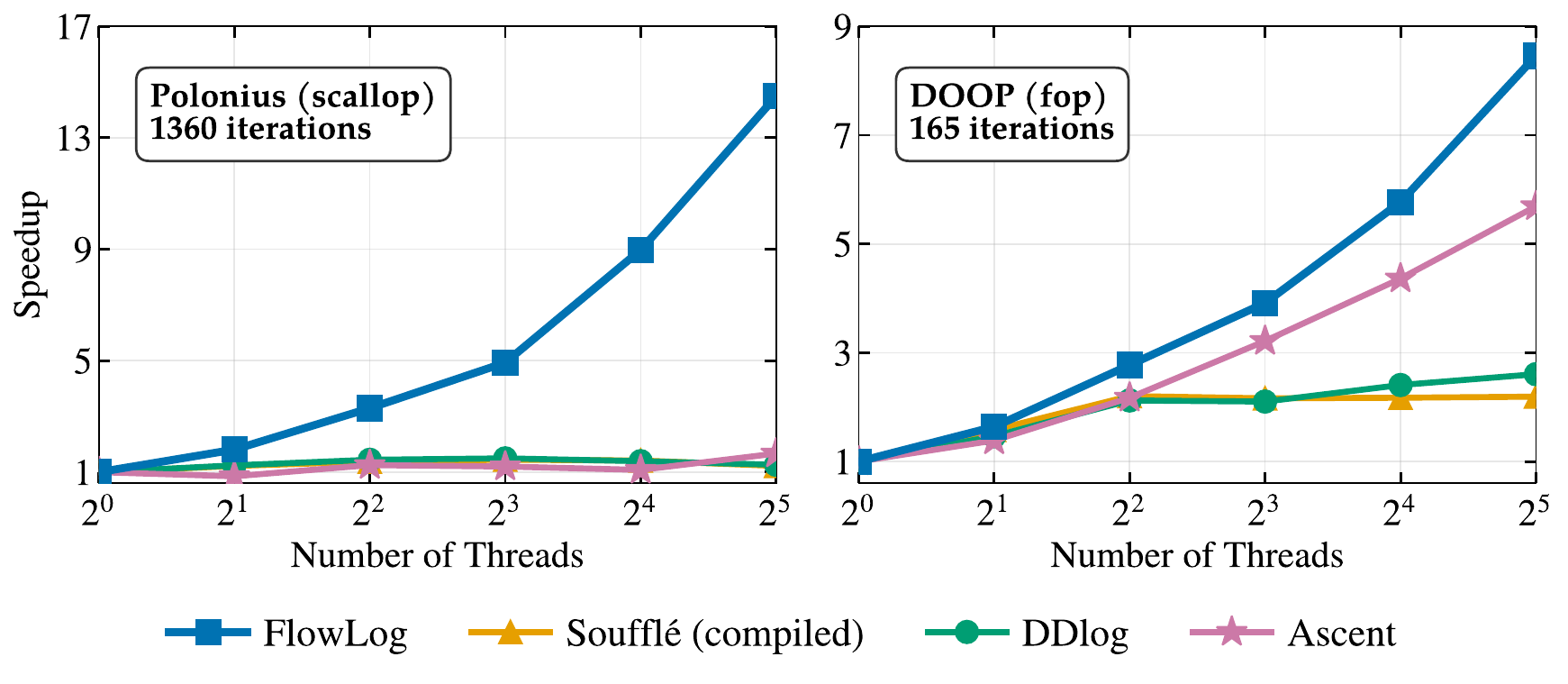}
  \caption{Scalability of all competing systems on two program-data pairs; each subplot shows speedups relative to the single-threaded run, up to $2^5$ threads. Legends give the program-data pair and iterations to converge.}
  \Description{Scalability plots showing speedup relative to single-threaded execution as worker count increases.}
  \label{fig:scalability}
\end{figure}

\subsubsection{Scalability} Static analyses converge through a fixpoint of many small iterations, each doing little work.  This makes parallelism hard: coordinating threads and exchanging data at every iteration can cost more than the iteration computes.  Figure~\ref{fig:scalability} shows the effect as we scale to 32 threads, where \souffle, \ddlog, and Ascent barely move past a single thread on the highly iterative \polonius workload.  \flowlog avoids this because it compiles to \textsf{DD}, whose asynchronous, data-parallel runtime keeps every worker busy across these lightweight iterations rather than stalling at each one, so it scales the most consistently, reaching $15\times$ on \polonius and $8\times$ on \doop.

%% file: sections/extensibility.tex

\section{Extensible Workflow}
\label{sec:ext}

\flowlog's relational \ir separates rule-level logic from the \textsf{DD} runtime, enabling extensions to execution, instrumentation, and semantics without re-engineering the core.

\begin{figure*}[t]
  \centering
  \begin{minipage}[t]{0.275\textwidth}
    \vspace{0pt}
    \begin{subfigure}[t]{\linewidth}
      \vspace{0pt}
      \centering
      \includegraphics[width=\linewidth]{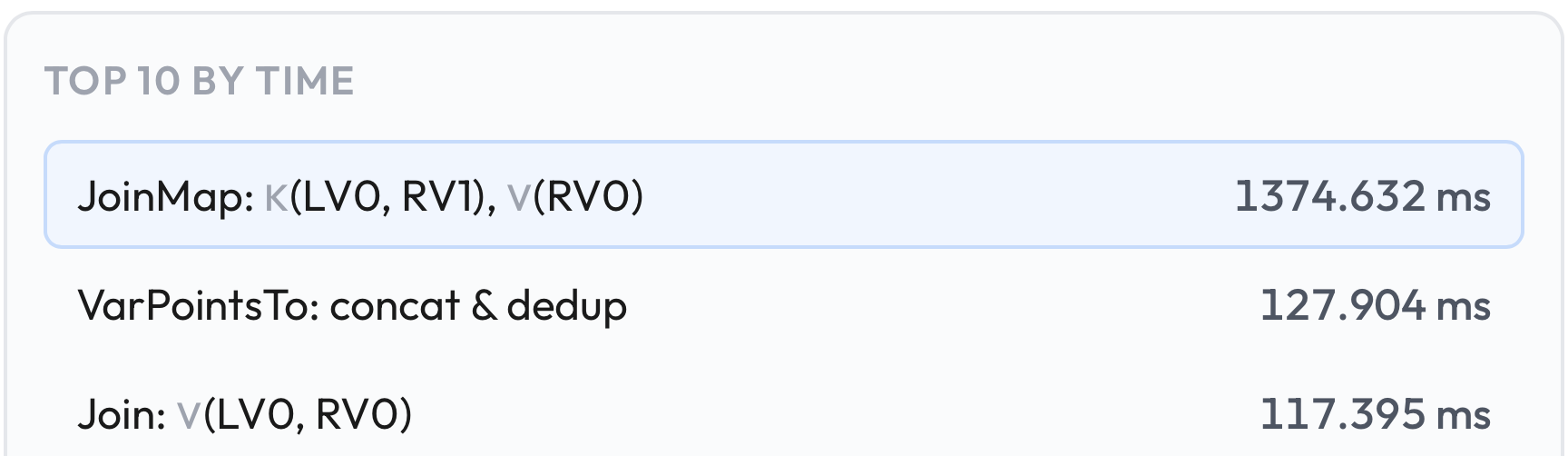}
      \caption{Operators ranked by active time.}
      \label{fig:profile-top}
    \end{subfigure}

    \smallskip

    \begin{subfigure}[t]{\linewidth}
      \vspace{0pt}
      \centering
      \includegraphics[width=\linewidth]{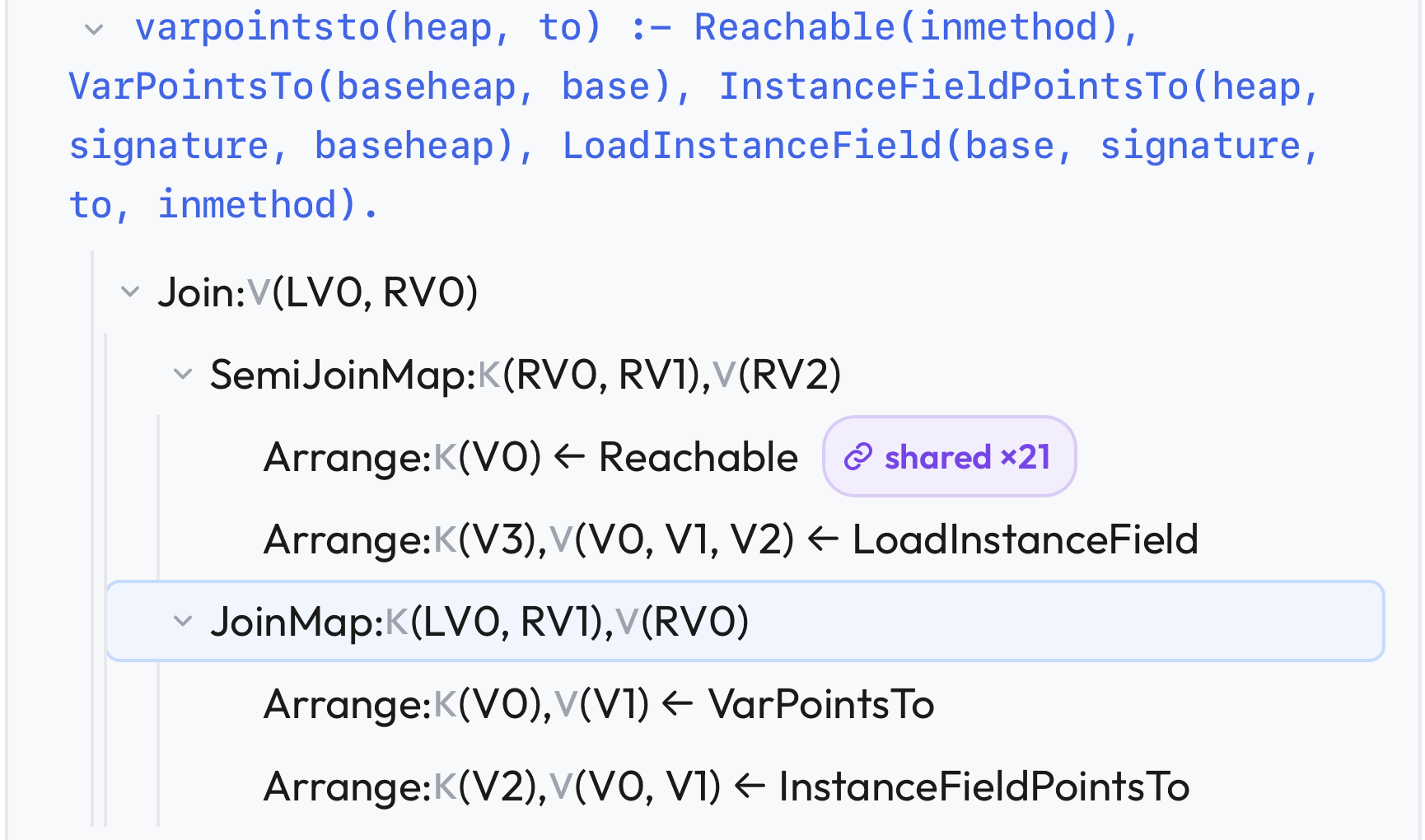}
      \caption{Rule plan with selected node.}
      \label{fig:profile-plan}
    \end{subfigure}
  \end{minipage}
  \hfill
  \begin{subfigure}[t]{0.705\textwidth}
    \vspace{0pt}
    \centering
    \includegraphics[width=\linewidth]{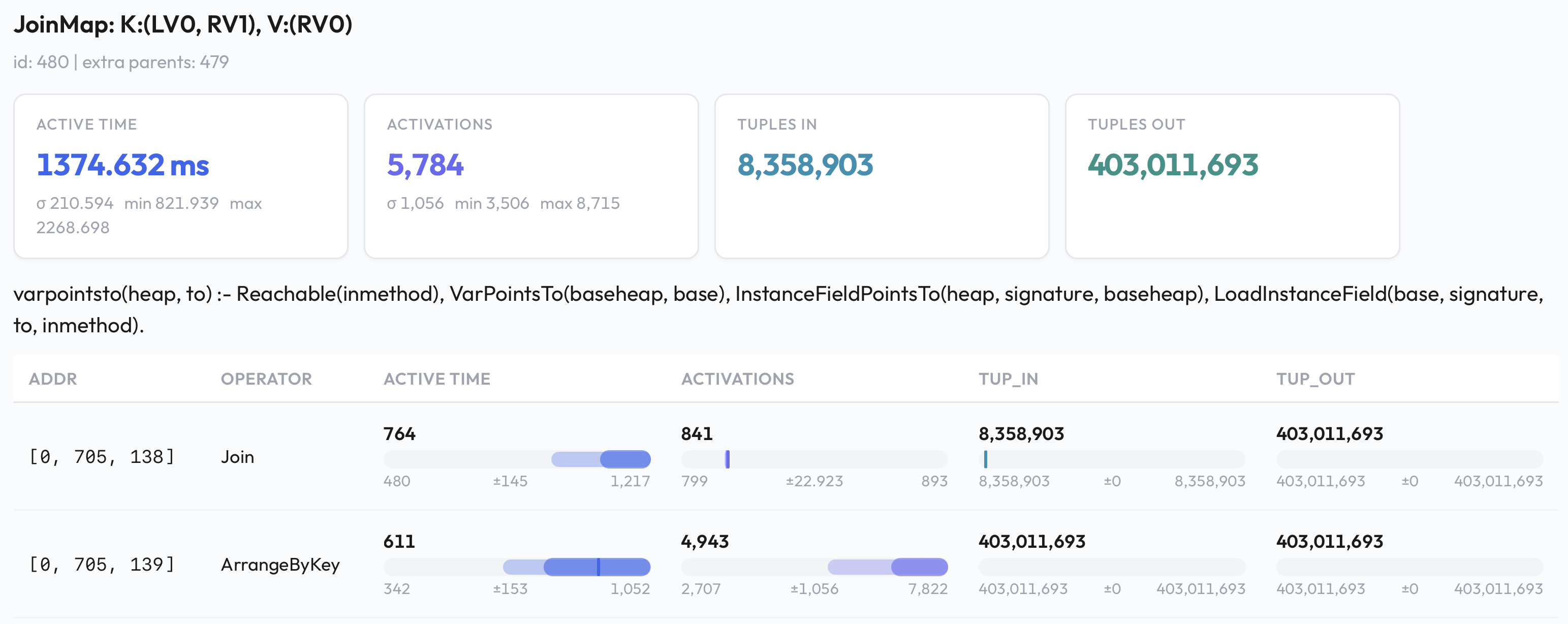}
    \caption{Selected-node runtime profile.}
    \label{fig:profile-node}
  \end{subfigure}
  \caption{\flowlog profiler views for diagnosing the poor join order in the \doop points-to rule from Example~\ref{ex:doop}.}
  \Description{Three FlowLog profiler views for the doop field-load rule under a forced poor join order. The time ranking selects a dominant JoinMap. The matching rule-plan node is highlighted, and its detail view reports 1.37 seconds of active time, 5,784 activations, 8.36 million input tuples, 403 million output tuples, and the underlying Join and ArrangeByKey operators.}
  \label{fig:profile}
\end{figure*}

\subsection{Incremental Maintenance}

To run incrementally, users add \texttt{--mode datalog-inc} (Table~\ref{tab:flags}) when compiling; this produces an executable with an interactive shell (Table~\ref{tab:inc-shell}). 

\begin{lstlisting}[numbers=none, basicstyle=\scriptsize\ttfamily, columns=fullflexible]
$ flowlog doop.dl -o doop-inc -F ./facts -D ./out --mode datalog-inc
\end{lstlisting}

\begin{table}[!htbp]
\centering
\caption{Interactive shell commands for incremental mode.}
\label{tab:inc-shell}
\footnotesize
\setlength{\tabcolsep}{4pt}
\renewcommand{\arraystretch}{1.08}
\begin{tabular}{@{}p{0.40\linewidth}p{0.55\linewidth}@{}}
\toprule
\textrm{\textbf{Command}} & \textbf{Description}\\
\midrule
\texttt{begin}
  & Begin an update round.\\
\texttt{put} \argrel{rel} \argtuple{tuple} \argdiff{diff}
  & Stage one tuple update.\\
\texttt{file} \argrel{rel} \argtuple{path} \argdiff{diff}
  & Stage updates from a file.\\
\texttt{commit}
  & Apply staged updates at one logical time.\\
\bottomrule
\end{tabular}
\end{table}

\begin{sloppypar}
We compile Example~\ref{ex:doop} in incremental mode and stage updates at two timestamps on the \texttt{tomcat} input: at \texttt{[t=0]} we load the base facts; at \texttt{[t=1]} we delete one \argrel{LoadInstanceField} fact, and \flowlog incrementally retracts the single \argrel{VarPointsTo} fact that depended on it. We also support batch updates through the \texttt{file} command.
\end{sloppypar}

\begin{lstlisting}[style=FlowLogShellAuto]
(@\textcolor[rgb]{0,0.5,0}{\ttfamily // t=0: load base facts and run the analysis}@)
[t=0] >> begin
[t=0] >> file (@\argrel{LoadInstanceField}@) (@\argtuple{load.csv}@) (@\argdiff{+1}@)
(@\textcolor[rgb]{0,0.5,0}{\ttfamily // ... load other EDB relations}@)
[t=0] >> commit
(@\textcolor[rgb]{0,0.5,0}{\ttfamily // tuples omitted; report only the VarPointsTo size}@)
(@\shout{[t=0] VarPointsTo size=4,986,481}@)
(@\textcolor[rgb]{0,0.5,0}{\ttfamily // t=1: remove one load from ParameterParser.chars}@)
[t=1] >> begin
[t=1] >> put (@\argrel{LoadInstanceField}@) (@\argtuple{ParameterParser.chars -> getToken/\$stack44}@) (@\argdiff{-1}@)
[t=1] >> commit
(@\shout{[t=1] VarPointsTo data=(char[]@String.toCharArray, getToken/\$stack44) diff=-1}@)
\end{lstlisting}

Applying this retraction takes about \SI{7}{\milli\second}, versus \SI{2.23}{\second} for full recomputation ($320\times$ faster), because \flowlog propagates only the affected deltas.

\subsection{Profiler-Guided Diagnosis}

\begin{sloppypar}
A poor join order can cripple a recursive rule, while choosing a good one from the rule text alone usually relies on expert intuition, trial and tuning.  \flowlog's profiler and visualizer expose per-operator cost, so users can localize bottlenecks without reading the generated dataflow code.  Users can set the order with a \texttt{.plan} directive over rule atoms.  In Example~\ref{ex:doop}, the three core atoms are \texttt{VarPointsTo}~($V$), \texttt{InstanceFieldPointsTo}~($I$), and \texttt{LoadInstanceField}~($L$); \texttt{Reachable} is always pushed down first as a semijoin (Section~\ref{sec:overview}), and \flowlog's default plan starts the joins from $L$, which is a good order.  To make the demonstration reproducible, we deliberately pin the bad order $(V \bowtie I) \bowtie L$ via the commented \texttt{.plan} directive. In normal use, users rarely know a good order in advance; they observe slow runs, then use the profiler and rule-level visualization to find expensive operators and compare plans.
\end{sloppypar}

Compiling with \texttt{-P} taps \textsf{DD}'s hooks in the generated dataflow to record per-operator metrics, and running the binary writes them to a log directory. \texttt{flowlog-visualizer}\footnote{\url{https://github.com/flowlog-rs/flowlog/tree/main/flowlog-visualizer}} renders these logs into a standalone HTML report. Figure~\ref{fig:profile} shows three parts of the profiler UI. Figure~\ref{fig:profile-top} ranks operators by active time; Figure~\ref{fig:profile-plan} shows the rule plan with its selected node; and Figure~\ref{fig:profile-node} shows the node's runtime profile. The runtime profile includes total input and output tuples, the underlying physical operators, and distributions of active time and activations across workers. It summarizes these distributions with their mean, standard deviation, minimum, and maximum, making load imbalance visible. Beyond the time ranking shown here, the UI provides top-10 lists for memory use and tuple blowup, as well as an interactive dataflow graph.

\begin{lstlisting}[numbers=none, basicstyle=\scriptsize\ttfamily, columns=fullflexible]
$ flowlog doop.dl -o doop-prof -F ./facts -D ./out -P
\end{lstlisting}

The profiler reports fused physical joins as \texttt{Join-Map} operators. A single \texttt{Join-Map} dominates the time ranking. Selecting it highlights the matching node in the rule plan (Figure~\ref{fig:profile-plan}), revealing the executed order $(V \bowtie I) \bowtie L$. Its Nodes view makes the bottleneck concrete: the operator processes $8.36$ million input tuples and emits $403$ million, with the physical \texttt{Join} and \texttt{ArrangeByKey} stages accounting for its active time (Figure~\ref{fig:profile-node}). The user closes the loop by deleting the \texttt{.plan} directive and restoring the default good order; recompiling cuts runtime from \SI{6.5}{\second} to \SI{2.2}{\second} on \texttt{tomcat} (\SI{220}{\second} to \SI{9.8}{\second} on \texttt{fop}).

\subsection{Extended Semantics}

Because \flowlog lowers programs to \textsf{DD} dataflows, it can support controlled extensions beyond standard \Datalog, including branching, loop control, and non-monotonic updates. We illustrate the non-monotonic case with k-core (Example~\ref{ex:k-core}), which standard \Datalog cannot express because derived facts are never retracted. In \flowlog, a \texttt{fixpoint} block re-runs its rules until stable, and \texttt{.iterative} marks relations allowed to shrink across rounds. The example keeps these semantics in the same compiled pipeline as ordinary \Datalog rules; the same mechanism can also capture equality-saturation-style rewriting in the spirit of egglog~\cite{egglog}. Full details appear in the online tutorial.\footnote{\url{https://www.flowlog-rs.com/tutorial/language/extended-semantics}}

\begin{example}[K-Core Computation]\label{ex:k-core}
Computing the $k$-core repeatedly removes vertices whose current degree is below $k$ until all remaining vertices have at least $k$ neighbors.
\begin{lstlisting}[style=DatalogStyle, numbers=none, basicstyle=\scriptsize\ttfamily]
// the block re-evaluates until it reaches a fixpoint
fixpoint {
    // .iterative marks relations that are non-monotonic
    .iterative active_edge
    .iterative degree
    active_edge(x, y) :- edge(x, y), !removed(x), !removed(y).
    degree(x, count(y)) :- active_edge(x, y).
    removed(x) :- degree(x, d), d < k.
}
\end{lstlisting}
\end{example}

%% file: sections/tool-availability.tex

\section{Tool Availability}
\label{sec:availability}

\begin{sloppypar}
\textbf{Tool:} \url{https://github.com/flowlog-rs/flowlog}.  \textbf{Docs:} \url{https://www.flowlog-rs.com/tutorial/intro/}.  Docs cover installation, command-line use, example analyses, incremental updates, profiling, and extended semantics.  \textbf{Video:} \url{https://youtu.be/DNU1Uzt47MI}.  \textbf{Archive:} \url{https://doi.org/10.5281/zenodo.20815412}.
\end{sloppypar}